\documentclass[twocolumn]{article}
\DeclareUnicodeCharacter{0308}{\"{}}
\usepackage[a4paper,top=2cm,bottom=2cm,left=1cm,right=1cm,marginparwidth=1.75cm]{geometry}
\usepackage{graphicx} % Required for inserting images
\usepackage{amsmath}
\usepackage{comment}
\usepackage[numbers,sort&compress]{natbib}
\usepackage{color}
\usepackage{booktabs}
\usepackage{braket}
\usepackage[usenames, dvipsnames]{xcolor}
\usepackage{bm}% bold math
\usepackage{hyperref}% add hypertext capabilities
\hypersetup{
    colorlinks=true,
    linkcolor=blue,     
    urlcolor=blue,
    citecolor=blue
}

\usepackage{authblk}
\usepackage{marvosym}  % For the email symbol
\usepackage[font=small]{caption} 
\begin{document}
\title{Real-time dynamics of the two-step charge-density-wave transition in bulk 1T-TaS$_2$}

\author[1,$\ast$]{Daeheon Kim}
\author[2,$\ast$]{Jinseok Oh}
\author[1]{Kahyeon Koh}
\author[1]{Yejun Cho}
\author[3,4]{Angel Rubio}
\author[2,$\dagger$]{Noejung Park}
\author[1,3,$\ddagger$]{Dongbin Shin}

\affil[1]{Department of Physics and Photon Science, Gwangju Institute of Science and Technology (GIST), Gwangju 61005, Republic of Korea}
\affil[2]{Department of Physics, Ulsan National Institute of Science and Technology (UNIST), UNIST-gil 50, Ulsan 44919, Republic of Korea}
\affil[3]{Max Planck Institute for the Structure and Dynamics of Matter and Center for Free-Electron Laser Science, Luruper Chaussee 149, 22761 Hamburg, Germany}
\affil[4]{Initiative for Computational Catalysis (ICC), The Flatiron Institute, 162 Fifth Avenue, New York, NY 10010, USA}

%\date{}
\maketitle
\renewcommand\thefootnote{\fnsymbol{footnote}}
\footnotetext[1]{These authors contributed equally.}
\footnotetext[2]{noejung@unist.ac.kr}
\footnotetext[3]{dshin@gist.ac.kr}
\renewcommand\thefootnote{\arabic{footnote}}

\section*{Abstract}
{ The charge-density wave (CDW) of bulk 1T-TaS$_2$ is built from Star-of-David (SoD) clusters tiling a $\sqrt{13} \times \sqrt{13}$ superlattice, and it melts through a two-step sequence accompanied by order-of-magnitude changes in resistivity.
Whether these steps proceed by collapse of the SoD amplitude or by rearrangement of the SoD lattice has remained unresolved, because the relevant dynamics occur on length and time scales beyond the reach of ab initio molecular dynamics. 
Here we follow the CDW transitions in real time using a machine-learning force field trained on first-principles data, giving access to 1404-atom supercells over 5 ns. The two steps are mechanistically distinct. 
Above 200 K, SoD clusters translate coherently by transiently dissolving and re-forming about shifted centers, a deformation–formation process that preserves the local SoD amplitude while randomizing the interlayer stacking order and nucleating domain walls. 
Only near 350 K does the SoD distortion itself collapse. 
These results provide microscopic support for the recently proposed two-step model of the CDW transition and offer a framework for interpreting light-induced hidden phases and cavity-modified transition temperatures in 1T-TaS$_2$.}

\section*{Introduction}
Charge density wave (CDW) states of 1T-TaS$_2$ have attracted wide interest for their tunable conditions, such as temperature, pressure, doping, optical pumping, and optical cavity~\cite{Sipos2008,Martino2020,yu_gate-tunable_2015,jarc2023cavity}.
Bulk 1T-TaS$_2$ undergoes two distinct phase transitions, resulting in substantial variations in electrical resistivity with temperature or doping conditions.
Thermodynamically, the transition between the commensurate CDW (CCDW) and near-commensurate CDW (NCCDW) occurs at $200$~K, and that between the NCCDW and incommensurate CDW (ICCDW) occurs at $350$~K~\cite{Sipos2008,Martino2020}.
Extensive scanning tunneling microscopy (STM) measurements highlight the importance of ionic distortions that modify electronic structure via strong electron-phonon interactions~\cite{fujii_electronic_2018,park_emergent_2019,gerasimenko_intertwined_2019,Ma2016}.
At the CCDW phase, for example, a Mott insulating CDW state, which is characterized by a compact Star-of-David (SoD) lattice distortion pattern of Ta atoms, results in high electrical resistivity (see inset of Fig. 1(b)).
At the NCCDW phase, the electrical resistivity becomes lower than that of the CCDW phase, and domain wall formations are observed through STM measurement~\cite{fujii_electronic_2018,park_emergent_2019,gerasimenko_intertwined_2019,Ma2016}.
At the ICCDW phase, complete deformation of the SoD structure, which induces the lowest electrical resistivity, is expected at high thermal energies above $350$~K.
These experimental observations are believed to originate from strong electron-phonon interactions.
In this context, several approaches for controlling the CDW phases of 1T-TaS$_2$ have been proposed, including the light-induced hidden phase and cavity-induced CDW phase transitions~\cite{stojchevska_ultrafast_2014,jarc2023cavity}.

Recent experimental and theoretical studies examined the microscopic mechanism behind the CCDW-NCCDW phase transition.
Especially, STM measurements reveal the formation of domain walls, leading to the NCCDW phase from the commensurately formed SoD structures in the CCDW phase~\cite{gerasimenko_intertwined_2019,de_la_torre_dynamic_2025}.
This observation led to the initial conjecture that the CCDW--NCCDW phase
transition, and the accompanying change in electrical conductivity, is driven
by the partial deformation of the SoD structure constituting the domain
walls~\cite{Sipos2008,gerasimenko_intertwined_2019}.
However, recent experiments demonstrated that this phase transition is absent in thin films ($d <$ 7 nm), as shown in Fig.~1(a)~\cite{Yoshida2017}.
This discrepancy indicates that in-plane structure modification does not solely induce the CCDW-NCCDW phase transition.
Recent theoretical studies also suggest that the CCDW phase originates from a specific stacking configuration known as AL stacking, which shows a non-negligible band gap (see Fig.~1(c))~\cite{Ritschel2018, Lee2019}.
Following these works, various stacking configurations and their corresponding band gaps are also experimentally identified~\cite{dong2023electronic, Butler2020}.
These recent observations indicate that the stacking order of SoD layers along the out-of-plane direction plays a crucial role in the CCDW-NCCDW phase transition.

\begin{figure}[t]
\includegraphics[width=0.48\textwidth]{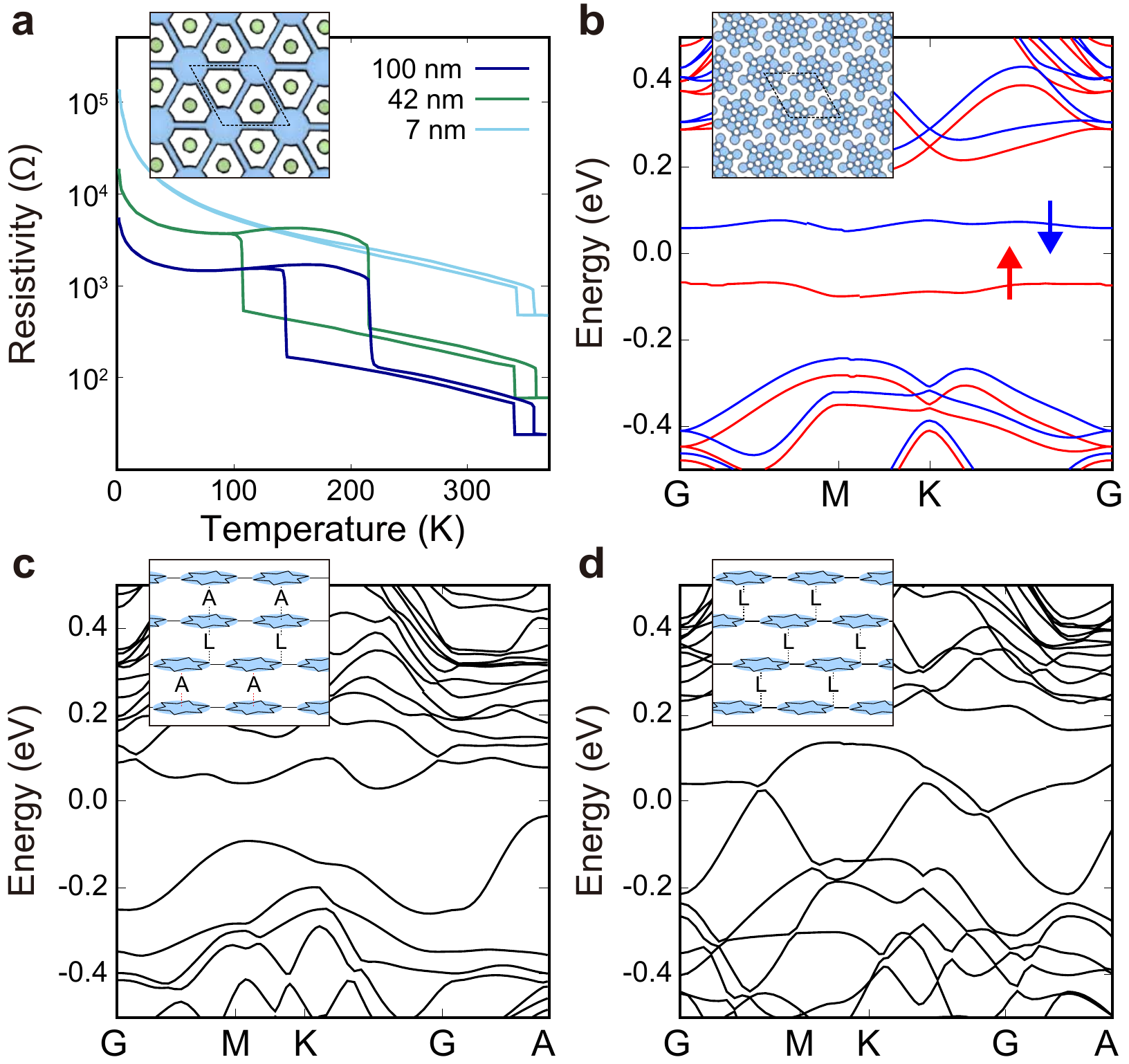}% Here is how to import EPS art
\caption{
\textbf{Temperature-dependent electrical resistivity and band structure of 1T-TaS$_2$ depending on the stacking configurations.}
 \textbf{a} Temperature-dependent electronic resistivity in thin film TaS$_2$. The data are adapted from those published in Ref.~\cite{yoshida_controlling_2014}.
\textbf{b-d} Band structure of 1T-TaS$_2$ in (b) monolayer, (c) AL stacking, and (d) L stacking configurations.
 Insets show schematic images for (a) 1T-TaS$_2$ in a primitive lattice without CDW formation, (b) the SoD lattice distortion in $\sqrt{13} \times \sqrt{13}$ lattices, (c) AL stacking, and (d) L stacking geometries.
 In the inset of (a), blue and green balls indicate Ta and S atoms, respectively.
 In (b), red and blue lines indicate the spin-up and spin-down bands.
}
\end{figure}

Our study investigates the microscopic mechanism behind the two-step CDW phase transition in 1T-TaS$_2$ through $ab~initio$ molecular dynamics (MD) simulations.
We explore the dynamics of SoD distortions to understand the thermal effects on SoD structures and their stacking configurations.
Utilizing extensive $ab~initio$-based MD simulations combined with machine learning force field (MLFF) techniques, which allow nanosecond simulations of bigger supercells containing up to 36 SoDs, we show that the formation of domain walls and randomization of SoD stacking can be achieved at high temperature ($T> 200$~K) via the deformation-formation process with coherent SoD breathing motions.
We further demonstrate that CDW deformation is directly linked to the NCCDW-ICCDW phase transition, lowering electrical resistance regardless of stacking configuration.
Our study reveals the microscopic dynamics mechanism underlying the two-step phase transitions in the 1T-TaS$_2$ system.

\section*{Results}
\subsection*{Stacking-dependent electronic structure in 1T-TaS$_2$}
The electronic structure of 1T-TaS$_2$ depends on the stacking configuration of SoDs along the out-of-plane direction.
At low temperatures, 1T-TaS$_2$ exhibits a compact SoD lattice distortion between Ta atoms in the $\sqrt{13} \times \sqrt{13}$ in-plane supercell, with CDW formation originating from the primitive cell (see the insets of Figs.~1(a) and~1(b)).
Because of the odd number of electrons in the unit cell of the monolayer 1T-TaS$_2$, its Mott insulating phase introduces a spin gap between spin-up and spin-down states near the Fermi level, as shown in Fig.~1(b).
Stacked 1T-TaS$_2$ layers along the out-of-plane direction introduce the band dispersion along the out-of-plane direction ($G-A$ line).
The AL stacking consists of alternating A-order and L-order stackings, as depicted in the inset of Fig.1(c).
In the A-order stacking, the central Ta atom is located directly above the corresponding Ta atom in the neighboring layer, whereas in the L-order stacking, the central Ta atom is positioned above a vertex of the SoD cluster in the neighboring layer.~\cite{Ritschel2018,Lee2019}.
Notably, this AL stacking is thought to be the ground state with the lowest DFT total energy among the various stacking configurations and provides a non-zero band gap.
Unlike AL stacking, the L stacking, as presented in Fig. 1(d), provides a metallic band structure originating from the higher out-of-plane hopping between CDW states~\cite{Ritschel2015}.
Based on these results, it is suggested that the reduced electrical resistivity at higher temperatures (T>$200$~K) is due to the emergence of the L-order stacking from the insulating ground-state AL stacking~\cite{Ritschel2018,Lee2019}.
However, the detailed microscopic mechanism underlying the stacking-configuration changes and domain-wall dynamics under thermal fluctuations has not been clarified.
A detailed description of the electronic structure with various stacking configurations and on-site Coulomb interactions~\cite{Shin2021} is provided in the Supplementary Section S1.

\begin{figure}[t]
\includegraphics[width=0.48\textwidth]{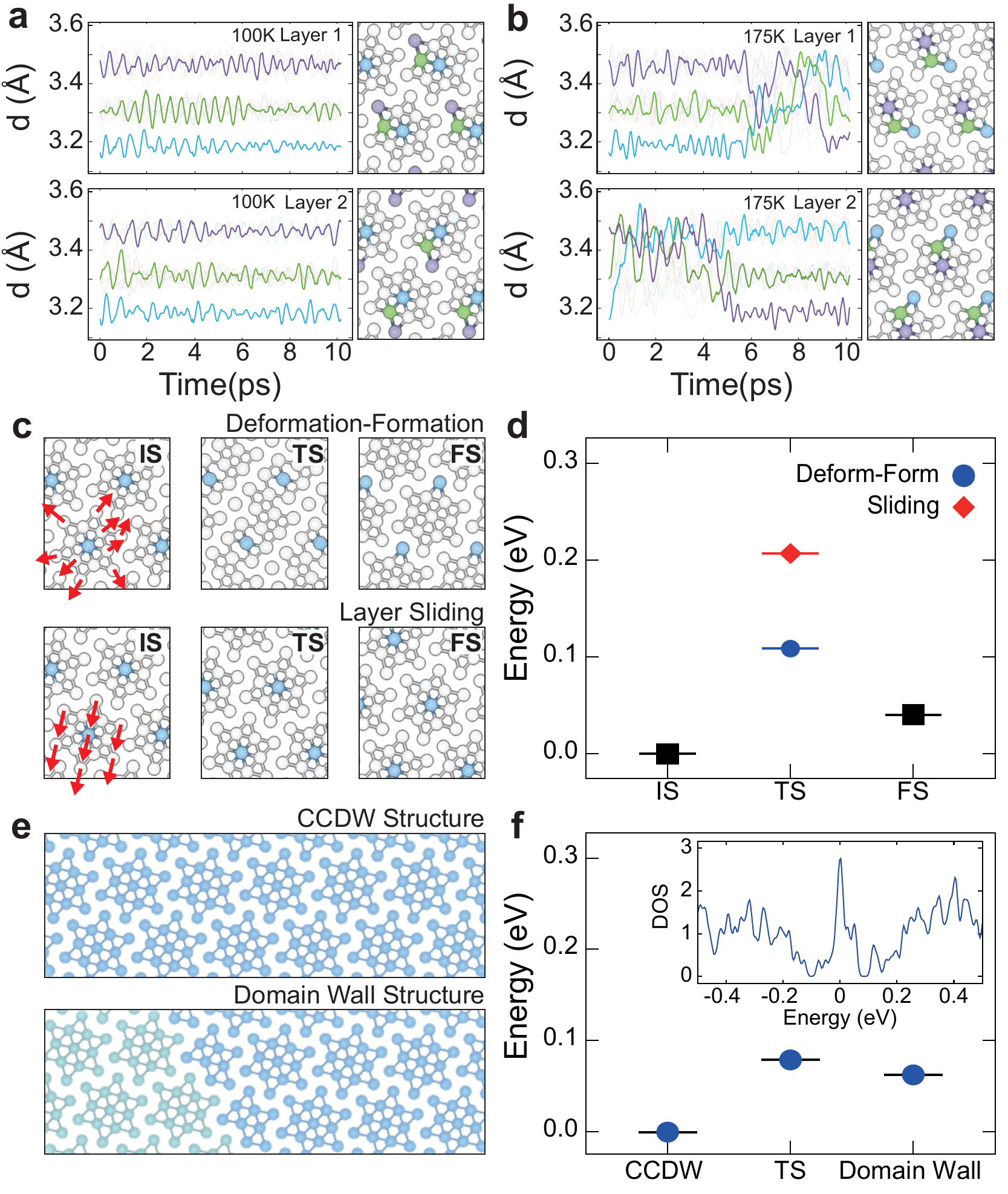}% Here is how to import EPS art
\caption{
\textbf{First-principles analysis for stacking order randomization and domain walls. }
\textbf{a-b} Time-profile of averaged nearest Ta-Ta distance ($d$) for each Ta atom during the $ab~initio$ molecular dynamics simulations with thermostat temperatures at (a) 100~K and (b) 175~K.
\textbf{c} Atomic geometries of SoD shifting via the deformation-formation process (upper) and CDW sliding (below).
\textbf{d} Transition barrier for the SoD shifting via the deformation-formation process and sliding.
\textbf{e} Atomic geometry of 1T-TaS$_2$ without (upper) and with an in-plane domain wall (below).
\textbf{f} Transition barrier for in-plane domain wall formation in $6\sqrt{13}\times \sqrt{13} $ monolayer 1T-TaS$_2$ lattice.
In the inset of (a), blue and green balls indicate Ta and S atoms, respectively.
In the inset of (c), red arrows indicate the displacement direction toward the transition state. 
The inset of (f) indicates the density of electronic states for the domain wall structure.
}
\end{figure}

\subsection*{Ab initio molecular dynamics reveal SoD shifting and domain wall formation}
To investigate the temperature-dependent dynamics in 1T-TaS$_2$, we perform AIMD simulations at $T = 100$~K and $T = 175$~K using an NVT ensemble.
As an initial condition, we employed the A-stacked 1T-TaS$_2$ in a $\sqrt{13} \times \sqrt{13} \times 2$ supercell.
The averaged distances between each Ta atom and its six nearest Ta neighbors are evaluated to characterize the thermal dynamics, as shown in Figs.~2(a) and 2(b).
Three distinct Ta sites are considered: the center, the inner vertex, and the outer vertex Ta atoms are depicted by sky blue, green, and purple balls, respectively.
At $T = 100$~K, there are no significant changes in the SoD geometry or stacking configuration for both the upper and lower layers.
On the other hand, the higher thermal energy ($T = 175$~K) introduces a dramatic change in the stacking configuration, as shown in Fig. 2(b).
In the upper layer (upper panel), the central Ta atom (sky blue) remains in place until around $6$~ps, after which its bond length increases.
Instead of this Ta atom (sky blue), the outer vertex Ta atom (purple) becomes the central atom after $9$~ps.
In the lower layer (lower panel), the Ta-Ta bond distances change abruptly until $4$~ps, and the outer vertex Ta atom (purple) becomes the central atom for SoD after $5$~ps.
This AIMD result indicates that sufficient thermal energy ($T>175$~K) can induce a change in the stacking order via the deformation-formation process.

We examine other possible motions that may appear at temperatures near the CCDW-NCCDW phase transition.
First, layer sliding can induce SoD shifting similar to the deformation-formation process.
A nudged elastic band calculation evaluates the transition barrier between the A stacking (initial state) and LC stacking (final state) geometries, as shown in Fig.~2(c).
The SoD shift via the deformation-formation process requires a $108$~meV transition barrier, whereas via layer sliding it requires a $207$~meV transition barrier (see Fig.~2(d)).
This indicates that SoD shifting through the deformation–formation mechanism is energetically more favorable than layer sliding~\cite{fujii_electronic_2018,park_emergent_2019,gerasimenko_intertwined_2019}.
As a second case, we consider the formation of a domain wall.
To investigate the transition barrier for the domain wall, we construct the $6\sqrt{13}\times\sqrt{13}$ superlattice containing a domain wall composed of partially deformed SoDs that separates two CCDW regions, as shown in the bottom panel of Fig. 2(e).
Through the nudged elastic band calculation, a $79$~meV transition barrier is obtained toward domain wall formation from the compactly formed SoD case, as shown in Fig.~2(f).
Notably, the non-zero reverse transition barrier ($16$~meV) indicates that this domain wall structure is metastable.
This result indicates that the partially deformed SoD with a domain wall can be metastable and involve the deformation-formation process.
Additional results on the time profile of the Ta-Ta distance at various temperatures are provided in Supplementary Section S2.

\begin{figure}[t]
\includegraphics[width=0.46\textwidth]{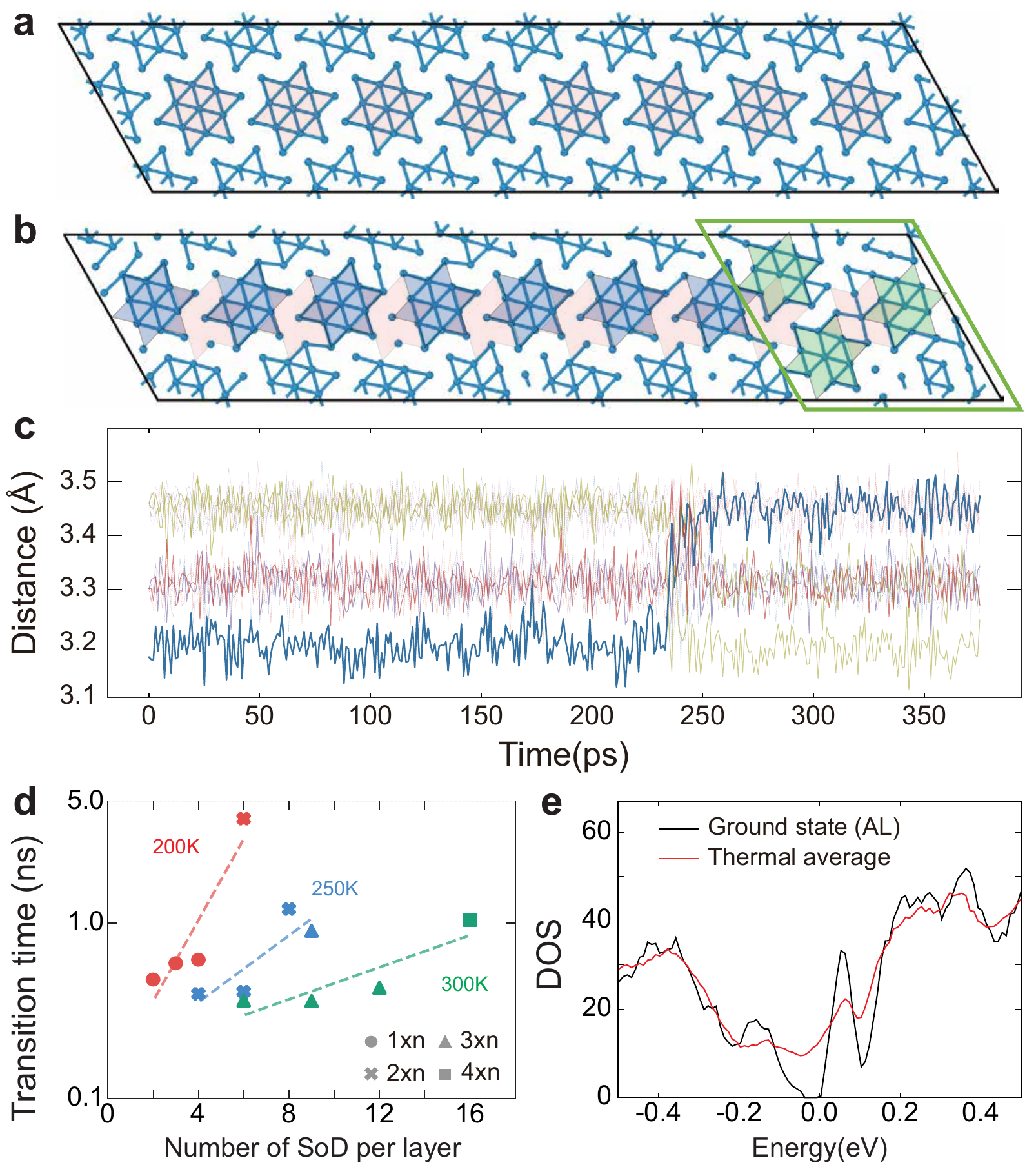}
\caption{
\textbf{Classical MD simulation with MLFF for the thermal dynamics in the extended superlattices.}
\textbf{a-b} Schematic geometry of $9\sqrt{13}\times 2\sqrt{13}\times 2$ super lattice with AL stacking for 1T-TaS$_2$ at (a) its ground geometry and (b) thermally excited geometry with domain wall formation at $0.2$ ns with thermostat at $200$~K.
\textbf{c} Time profile of the averaged nearest Ta-Ta distance in the MD with a thermostat at $200$ K.
\textbf{d} Transition time for SoD shifting depending on the size of the superlattice and temperature.
\textbf{e} The DOS of AL stacked 1T-TaS$_2$ and averaged DOS for various stacking configurations.
}
\end{figure}

\subsection*{Stacking-order randomization induced by coherent motion in the extended superlattice}

The coherent SoD shifting induces both stacking order randomization and the formation of a domain wall.
To explore the microscopic dynamics with many CDW sites, we perform classical MD simulations with MLFF in the
$9\sqrt{13}\times 2\sqrt{13} \times 2$ extended AL stacking configuration.
At $t=0$, the 36 SoDs are commensurately aligned, as shown in Fig.~3a.
When the $200$~K thermostat is considered in this extended lattice, the SoD shifting and formation of the domain wall occur, as shown in Fig.~3b.
Similar to the AIMD result presented above, the time profiles of the Ta–Ta distance reveal SoD shifts (see Fig.~3c).
The center Ta atom (blue) for the SoD structure initially exhibits a Ta–Ta distance of $3.2$~\AA.
Then, this distance dramatically increases to $3.45$~\AA~at 230~ps, with the vertex Ta atom (yellow) subsequently becoming the new center.
This observation confirms that SoD shifting is achieved in the extended lattice with a substantially elongated transition time.
Notably, this SoD shifting also introduces a domain wall formation, as indicated in the green box of Fig.~3b.

The SoD transition time depends on the size of the coherent SoD motions.
Unlike the previous $\sqrt{13}\times\sqrt{13}$ lattice (two SoDs, $\tau_s^{1\times1}=4$~ps), the extended lattice exhibits a much longer SoD shifting time (36 SoDs, $\tau_s^{9\times2}=230$~ps).
This cell-size dependency is attributable to the coherent breathing motion of the CDW during the deformation-formation process.
In the $\sqrt{13}\times\sqrt{13}$ lattice, the periodic boundary condition enforces coherent motion of all CDWs, maximizing the probability of overcoming the barrier for collective SoD shifting.
On the other hand, the extended lattice grants the CDWs additional freedom in their phase of motion, thereby suppressing the efficiency of such collective SoD shifting.
For instance, only when a sufficient number of SoDs accidentally breathe in phase does the SoD shift occur, which eventually leads to domain wall formation.
To examine how this increased degree of freedom in CDW motion degrades the coherence and delays the transition, we performed MD simulations on $n\sqrt{13}\times m\sqrt{13}$ supercells ($n\times m$) starting from AL stacking at various temperatures.
As presented in the logarithmic plot in Fig.~3d, the transition time increases exponentially with lattice size and temperature.
Because the number of degrees of freedom of SoD motion in experiments far exceeds that of our simulations with small periodic supercells, ultrafast SoD shifting ($\tau_s<0.1$~ns) might be hard to observe in the STM images.

\subsection*{Origin of the metallization at the CCDW–NCCDW transition}

The randomization of the stacking order in 1T-TaS$_2$ is a critical factor for the metallization in the NCCDW phase.
Recent STM measurements of surface states revealed the emergence of domain walls featuring partially deformed SoDs in the NCCDW phase, suggesting that these structures might contribute to the metallic electronic properties~\cite{gerasimenko_intertwined_2019,wang2020band,wang2024dualistic}.
If in-plane domain wall structures were the primary driver of metallicity in the NCCDW phase, the CCDW-NCCDW transition should occur in both monolayer and thin-film samples, irrespective of thickness~\cite{yoshida_controlling_2014}.
However, the CCDW-NCCDW transition is absent in the thin-film limit, as shown in Fig.~1a~\cite{yoshida_controlling_2014}.
Furthermore, domains with sizes of a few tens of nanometers exhibit distinct electronic characters—either metallic or insulating—indicating that the electronic properties are not significantly affected by the in-plane domain-wall structure~\cite{Butler2020,zhang_visualizing_2022,yang_thickness_2023}.
These results suggest that a comprehensive combination of stacking configurations, including the insulating AL-stacked ground state, metallic L stacking, domain-wall structures, and other stacking configurations, cooperatively determines the electrical resistivity of 1T-TaS$_2$.
For example, while commensurate AL stacking yields an insulating DOS, the averaged ($T=200$~K) DOS over various stacking configurations is metallic, as shown in Fig.~3e.
The DOS associated with the domain-wall structure is also metallic but exhibits a sharp peak at the Fermi level, as shown in the inset of Fig.~2f.
This sharp peak, however, is not observed in experimental $dI/dV$ spectra of the metallic phase of 1T-TaS$_2$ over a wide range of samples~\cite {zhang2022reconciling,wu2022effect,Butler2020}.
Because thicker 1T-TaS$_2$ can have a dispersive band structure along the out-of-plane direction ($\Gamma$-A), randomization of the stacking configuration can provide a metallic electronic structure. 
This result indicates that the thickness dependence of the low-temperature electrical resistivity originates from the higher probability of metallic stacking configurations in thicker samples~\cite{yoshida_controlling_2014}.
In contrast, thin films retain an insulating band structure owing to the suppressed out-of-plane band dispersion~\cite{park_stacking_2023}, consistent with the fact that the CCDW-NCCDW transition is not observed in the thin-film limit (see also the SI for the band structure of few-layer 1T-TaS$_2$).
These results suggest that the insulating AL stacking is the ground state of the CCDW phase, and that modified stacking configurations, rather than in-plane domain-wall formation alone, drive the metallicity in the NCCDW phase.

\begin{figure}[t]
\includegraphics[width=0.48\textwidth]{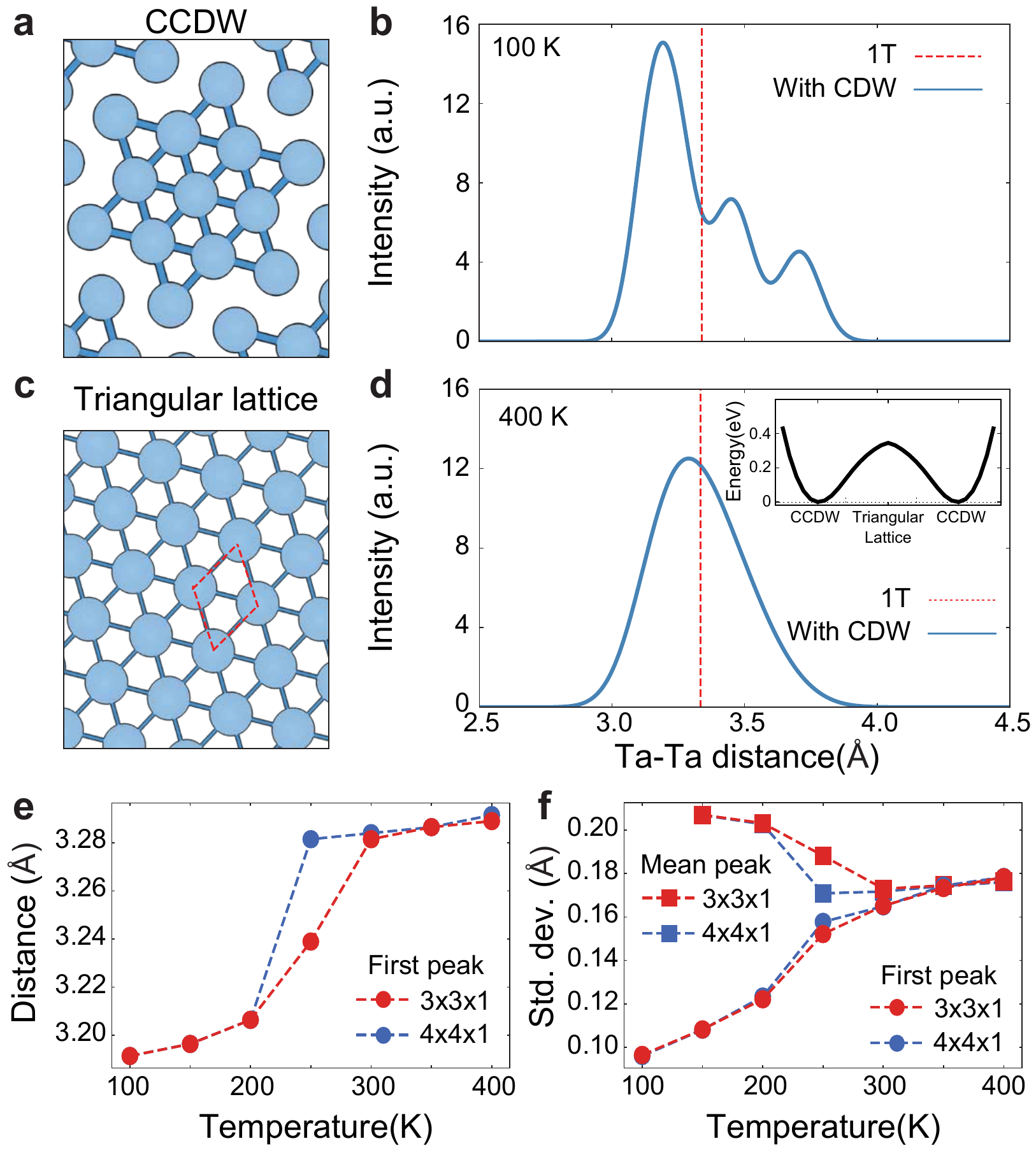}% Here is how to import EPS art
\caption{
\textbf{Deformation of SoD structure achieved by the higher thermal energy.}
\textbf{a} The atomic geometry of 1T-TaS$_2$ with SoD formation at ground-state geometry.
\textbf{b} Radial distribution function for Ta-Ta atomic positions at $100$~K.
\textbf{c} Triangular lattice of 1T-TaS$_2$ prior to the SoD formation.
\textbf{d} Radial distribution function for Ta-Ta atomic positions at $400$~K.
\textbf{e} Temperature dependence of the peak position of the radial distribution for nearest Ta-Ta distances.
\textbf{f} Temperature dependence of standard deviations of the first and mean RDF peaks.
Inset of (d) indicates potential energy surface along with and without SoD formation.
}
\end{figure}

\subsection*{Thermally driven SoD deformation and the NCCDW-ICCDW transition}

The NCCDW-ICCDW phase transition originates from the full deformation of SoD geometry.
The time-averaged radial distribution function (RDF) from the MD simulation is evaluated to analyze the atomic geometry at a given temperature.
The RDF provides separated peaks at low temperatures ($100$~K), indicative of commensurate SoD distortions, as shown in Figs.~4a and 4b.
This SoD distortion provides the highest $3.18$~\AA~peak originating from the Ta-Ta atomic distance between intra-SoD sites for the center Ta atom for SoD.
At high temperature ($400$~K), on the other hand, the RDF becomes broadened, and the peak point is shifted to $3.28$~\AA,~which corresponds to the Ta-Ta distance of the $1\times 1$ primitive lattice without SoD formation.
Notably, fully deformed SoD geometry can be achieved at a higher temperature to overcome the higher transition barrier ($0.35$~eV/SoD), as shown in the inset of Fig.~4d.
When the thermal energy is sufficiently high, the probability of complete SoD deformation becomes sufficiently large, enabling the NCCDW–ICCDW phase transition irrespective of the stacking configuration or sample thickness.
The RDFs exhibit distinct patterns depending on each CDW phase.
The position of the first RDF peak for the Ta-Ta distance gradually increases with temperature, as shown in Fig.~4e.
The Ta-Ta distance grows from $3.18$~\AA\ at low temperature to $3.28$~\AA\ at $250$~K, with the latter corresponding to the Ta--Ta distance in the symmetric triangular 1T-TaS$_2$ without SoD formation.
This result indicates that the mean Ta-Ta separation gradually increases through SoD shifts and thermal expansion.
On the other hand, the standard deviation of the Ta-Ta bond length distinguishes the NCCDW phase from the other phases.
As shown in Fig.~4f, the standard deviations of the Ta-Ta bond length obtained from the first peak and the mean peak are $0.096$~\AA\ and $0.209$~\AA\ at $T=100$~K, respectively.
These values reflect minimal thermal vibrations at low temperature and the broad distribution of Ta-Ta distances induced by SoD formation.
At $T=350$~K, the standard deviations of the two peaks merge, indicative of complete removal of SoDs owing to high thermal energy.
These statistics of the Ta dynamics provide clear indicators of the two-step CDW phase transition.
Additional analysis of the RDF function at various temperatures is provided in Supplementary Section S3.

\section*{Discussion}
We investigated the thermal dynamics of 1T-TaS$_2$ through extensive \textit{ab initio}-based simulations.
We performed AIMD and MLFF-based MD calculations at various temperatures to uncover the microscopic dynamics underlying the three phases of 1T-TaS$_2$.
The two steps are therefore mechanistically distinct: the first transition reorganizes the SoD lattice and the interlayer stacking via the deformation-formation process, whereas the second transition destroys the SoD amplitude itself.
Based on our results, the experimentally observed thickness dependence of the CCDW-NCCDW transition~\cite{yoshida_controlling_2014} can be attributed to stacking randomization.
In the bulk, randomized stacking configurations produce metallic band dispersion along the out-of-plane direction, whereas thin films retain an insulating band structure~\cite{park_stacking_2023}.
The temperature-dependent RDF analysis demonstrates that complete deformation of the SoD geometry occurs at the NCCDW-ICCDW phase transition.
Overall, our analysis provides a detailed picture of the microscopic mechanisms of the two-step phase transition in 1T-TaS$_2$, consistent with experimental observations~\cite{yoshida_controlling_2014,Butler2020,Lee2019,Ritschel2018}.
We suggest that our results also provide insight into light-induced CDW phase transitions.
For instance, the laser-driven hidden phase should retain the full intralayer SoD amplitude while losing interlayer correlations~\cite{stojchevska_ultrafast_2014}.
Therefore, the laser excitation is likely to trigger a coherent deformation-formation process and SoD shifting, randomizing the stacking order and thereby producing the metallic hidden phase.
Furthermore, the optical-cavity-induced CDW phase transition~\cite{jarc2023cavity} can be understood in terms of modified thermodynamics, such as a renormalized transition barrier for SoD shifting.

\section*{Methods}
\subsection*{DFT calculations}
We performed DFT calculations using the Quantum Espresso package \cite{Giannozzi2017} with a PBE-type functional~\cite{Perdew1996}.
The projector-augmented-wave method (PAW) is used for the atomic potentials, and the plane-wave basis set with a 60 Ry energy cutoff is used to describe the wavefunction of the 1T-TaS$_2$ system.
The Grimme-D3 correction is employed to account for van der Waals forces between TaS$_2$ layers~\cite{grimme_consistent_2010}.
The calculated lattice constant for bulk 1T-TaS$_2$ is $a=3.34$~\AA~and $c=5.9$~\AA~for primitive lattice, $a=3.34$~\AA~and $c=5.87$~\AA~for AL-stacking, and $a=3.34$~\AA~and $c=5.85$~\AA~for L-stacking, respectively.
For Brillouin zone sampling, a $3 \times 3 \times 6$ $\mathbf{k}$-point mesh was used for bulk 1T-TaS$_2$, while a $6 \times 6 \times 1$ $\mathbf{k}$-point mesh was used for the monolayer system.
The $3 \times 3 \times 2$ $\mathbf{k}$-point mesh was used in the nudged elastic band calculations with 10 intermediate images.

\subsection*{Machine-learning force field and large-scale molecular dynamics}
The LAMMPS package is employed~\cite{LAMMPS} to perform the large-scale classical MD with MLFF.
The MLFF was constructed using NequIP, an E(3)-equivariant graph neural network interatomic potential package~\cite{Batzner2022}.
For the training, more than $5000$ configurations sampled from AIMD trajectories at 100-500~K, generated with VASP~\cite{Kresse1996} in the NVT ensemble, were employed.
For the AIMD trajectories as training data, $2\sqrt{13}\times\sqrt{13}$ AL stacking 1T-TaS$_2$, a $5$~fs time step, and a thermostat with a velocity rescaling method are considered.
The $5$~fs time step and Nose-Hoover thermostat are used in MD simulations with MLFF at various temperatures.

\section*{Data availability}
The data that support the findings of this article are openly available at 10.6084/m9.figshare.33145793.

\section*{Code availability}
Quantum Espresso, LAMMPS, and NequIP are open-source packages available from their respective repositories; VASP is available under a commercial license from vasp.at. In-house scripts used for the MD analysis in this study are available from the corresponding authors upon request.

\providecommand{\noopsort}[1]{}\providecommand{\singleletter}[1]{#1}%

\section*{Acknowledgments}
\subsection*{Funding}
We acknowledge support by the Max Planck Institute New York City Center for Non-Equilibrium Quantum Phenomena, the Cluster of Excellence ``CUI: Advanced Imaging of Matter''--EXC 2056--Project ID 390715994, European Research Council (ERC-2024-SyG-101167294; UnMySt), and Grupos Consolidados (IT1453-22).
We were supported by the National Research Foundation of Korea (NRF) grants funded by the Korean government (MSIT) (No. RS-2023-00257666, RS-2023-00208825, RS-2024-00333664, and RS-2026-25508298).
The Flatiron Institute is a division of the Simons Foundation.

\section*{Competing Interests}
The authors declare no conflicts of interest.

\section*{Author Information}
\subsection*{Contributions}
D.S., N.P., and A.R. wrote the manuscript. 
D.K., K.K., Y.C., and J.O. performed the ab initio calculations and MD simulations under the supervision of N.P. and D.S.  
All authors discussed the results and contributed to the final paper.

 \subsection*{Corresponding authors}
Correspondence to Dongbin Shin or Noejung Park.

\end{document}